\documentclass[onecolumn,showpacs,preprintnumbers,amsmath,floatfix]{revtex4}
\usepackage{graphicx}
\usepackage{dcolumn}
\usepackage{bm}

\begin{document}

\title{Simulation of the radio signal from ultrahigh energy neutrino-initiated showers}

\author{Shahid Hussain}
 \email{vacuum@ku.edu}
\author{Douglas W. McKay}%
 \email{mckay@kuark.phsx.ku.edu}
\affiliation{Department of Physics \& Astronomy 
University of Kansas, Lawrence, KS 66045.}

\date{\today}

\begin{abstract}
PeV neutrinos produce particle showers when they interact with the atomic nuclei in ice. 
We briefly describe characteristics of these showers and the radio Cherenkov signal 
produced by the showers.  We study pulses from electromagnetic (em), hadronic, 
and combined em-hadronic showers and propose extrapolations to EeV energies.
\end{abstract}
\maketitle\section{Introduction}

Ultrahigh energy (UHE) neutrinos have energies of a PeV and higher. Although
these neutrinos have not been observed, a number of models predict their
existence. Another reason to believe in UHE neutrinos is
the existence of UHE cosmic rays which have been observed by a
number of detectors. Any models that account for these cosmic rays also
predict existence of UHE neutrinos. Ultrahigh energy neutrino detection will 
have a great impact on fundamental physics and astrophysics. Because of their 
ultrahigh energies, they will help us explore fundamental interactions at energies 
well beyond the
reach of man-made accelerators. Also, being neutral and presumably weekly
interacting, these neutrinos, unlike charged particles, will bring us
directional information about the source. Being weakly interacting they can help us 
explore regions of astrophysical objects which otherwise are opaque.

The electrons and hadrons from the neutrino-nucleon interactions will initiate particle 
showers as, unlike muons and taus, they dump their energy very
quickly in matter through electromagnetic and strong interactions. These showers produce
radio and optical radiation which can be detected\cite{zhs,amz,ketal1,ketal2} in detectors 
like RICE\cite{ketal1,ketal2} and ICECUBE\cite{icecube}. 
Radio ice Cherenkov experiment 
(RICE)\cite{ketal1,ketal2}, located at the south pole, can detect these showers 
from the radio signal they produce.

Here we give a brief introduction to the theory of radio emission from the
showers\cite{retal}. Particle showers in matter, unlike air showers, travel only a
few 10's of meters and dump all their energy in the target material. The
number of particles and the spacial size of the shower depends on the energy
of the primary particle. For ultrahigh energy primaries, the shower, at a
given time at shower maximum (which occurs at a few meters from the primary
vertex), has millions of particles located in a few cubic centimeters. 
As these particles are
moving faster than the speed of light in the given material, the charged
particles will emit Cherenkov radiation at all wavelengths. Whether this
radiation will be coherent depends on its wavelength, the net charge in the
shower, and the shower size (the volume in which most of the shower particles
 are contained at a given time). For a typical shower size (a few cubic
centimeters) one expects to get a coherent signal at Cherenkov angle due to
constructive interference of the radio radiation from different parts of the
shower. Hence one gets a
strong radio signal from the shower which makes the radio detection of the
showers (RICE) a very attractive technique as compared to the optical
detection (ICECUBE\cite{icecube}).

For radio detection of the showers, one needs a radio transparent material,
with a huge volume (to have detector effective volume large enough, of the
order of a cubic kilometer, to detect tiny fluxes of UHE neutrinos), in a
region with as low as possible background radio signal. The last requirement
makes a cold (to reduce background radio emission from atoms due to their
vibrations at finite temperature) and isolated area an ideal environment for
radio detection. Cold ice is remarkably transparent to radio transmission, with 
attenuation length of more than a kilometer.

RICE\cite{ketal1,ketal2}, founded in 1995, is located 100m below surface, 
above AMANDA\cite{hundertmark}, at the
South pole. The instrumented volume is (200m)$^{3}$ which consists of 20 radio
dipole antennas optimized to detect 0.5GHz radio signal. The effective
volume depends on signal strength which in turn depends on the energy of the
primary. For UHE showers, RICE effective volume is of the order of a km$^{3}$%
. RICE expects to upgrade its volume by an order of magnitude in
the coming years.

\section{ Shower simulations}

As mentioned above one expects to get a coherent radio signal from the
particle showers produced by UHE neutrinos. The dependence of this signal on
the distance between the shower and the observer is trivial: the signal goes
like 1/R, where R is the distance, with a weak exponential attenuation. 
We want to know, at a given neutrino energy:
How will this signal vary with angle (of the cone around shower axis) and
frequency (of the emitted radiation)? How will this signal vary for
different primary particles (from UHE neutrino interaction with nuclei)?
At ultrahigh energies, far above the reach of accelerators, 
one has to rely on numerical simulations of
the radio signal from the showers which contain millions of particles. 

To simulate the shower signal in ice, one propagates the shower
particles in ice and calculates the radio signal from them at every step
in space and time\cite{retal}. Here we extend the simulation to ultrahigh energies. In this 
case the shower contains millions of particles at the shower maximum. We have used PYTHIA
\cite{pyth} as an event generator software to produce an event vertex due to a
neutrino-nucleon interaction. Then we feed these particles to GEANT4\cite{geant}. 
assuming they are located at the same point (event vertex) in space and
time \footnote{For an expert reader, GEANT4.5.2 includes LPM effect in 
bremmstraulung only.
LPM effect in pair production has not been implemented yet in GEANT.}.

\section{Simulation results}

\begin{figure}
\includegraphics[width=3.5in,angle=0]{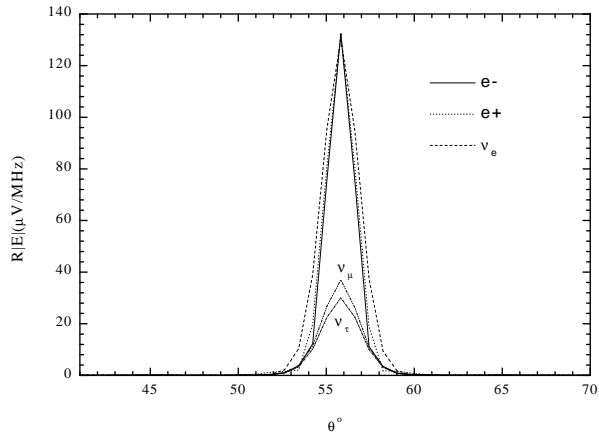}
\caption{\label{fig:fvsth}$R|E|$(where R is the distance from the shower to the observer 
and E is the electric field) vs $\theta$ (the angle of observation from the shower axis), 
at 1GHz (signal frequency), 
and E=1PeV (energy of the primary). 
The peak occurs at Cherenkov angle for ice. For neutrinos, the
elasticity value ($1-y$) is 0.65.}
\end{figure}

For CC interaction of the neutrinos, $\nu_{\mu}$ and $\nu_{\tau}$ are
not as efficient as $\nu_{e}$ in producing the signal. This is because the muons
and taus produced in a CC interaction take away some energy and do not
contribute to the shower (at these energies, typically they can go a long
way without decaying; also their electromagnetic (em) energy losses are very
small as compared to the electrons) while electrons will dump all their
energy into em energy which contributes to the signal. In Fig.~\ref{fig:fvsth} we
see the signal at a fixed frequency peaks at the Cherenkov angle in ice as expected. 

\begin{figure}
\includegraphics[width=3.0in,angle=-90]{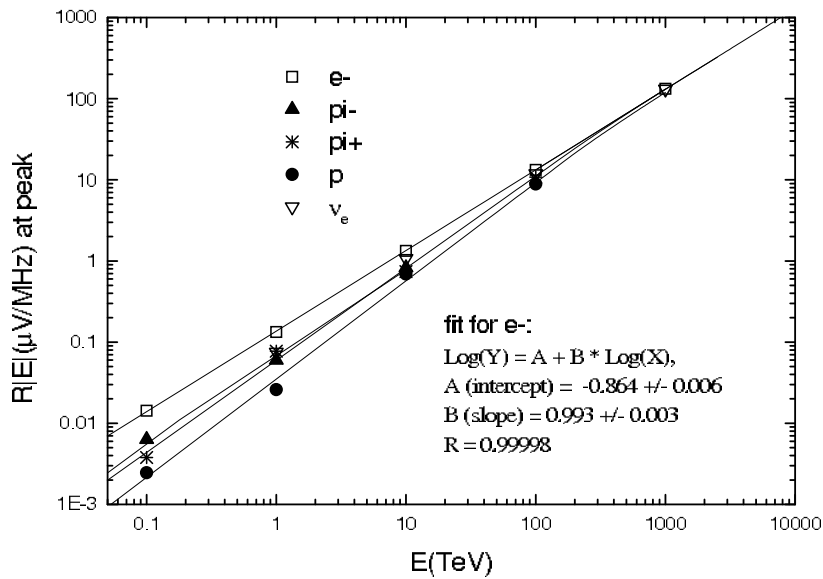}
\caption{\label{fig:fvse}$R|E|$ at $\theta _{c}$ and 1GHz vs energy for different particles.
The best fit for $e^{-}$ is also shown. For neutrinos, the
elasticity value ($1-y$) is 0.65.}
\end{figure}

Next we look at the energy dependence of the signal from em and hadronic showers.
Figure~\ref{fig:fvse} shows the signal as a function of energy
for different primaries. We see although hadrons, as compared to electrons,
are less efficient in producing the signal at lower energies, at higher
energies they are equally efficient. We extrapolate the signal for hadrons
above a 100TeV and for electrons and neutrinos above a PeV. We see the fit
for electrons is almost linear.
\section{Summary}

Ultrahigh energy neutrinos are a window to explore fundamental physics and
astrophysics. They have not been observed/identified yet. However, a number
of experiments have observed UHE cosmic rays and almost all the models that
account for these cosmic rays also predict the existence of UHE neutrinos.
These neutrinos can produce particle showers when they interact with matter.
These showers produce coherent radio signal while their optical signal is
incoherent. The strength of the radio signal depends on the net charge in
the shower which in turn depends on the number of particles in the shower
which increases with neutrino energy. This makes radio detection technique
very powerful at UHE energies (as the coherent radio signal is much larger
than the optical signal at these energies). RICE expects to detect the radio
signal from these UHE initiated neutrino showers. So far RICE has produced
strict upper bounds on some of the UHE neutrino flux models.

Simulation results show that hadronic showers, as compared to EM showers for
energies below a PeV, are not as efficient in producing radio signal.
However, as one goes to ultrahigh energies, hadronic and EM showers become
equally efficient in producing radio signals . The signal rises linearly
with energy at ultrahigh energies.

UHE neutrino detection situation is promising. AMANDA and RICE are taking
data. RICE expects upgrade in the coming years which will increase its
volume by an order of magnitude. ICECUBE and ANITA are funded and in
development stages. A number of other projects are underway.

\section*{Acknowledgments}
Shahid thanks to the LLWI (2004) organizers, the Graduate school of 
the University of Kansas, and the high energy group there for the travel funds.

\end{document}